\documentclass[fleqn,twoside]{article}
\usepackage{espcrc2}

\usepackage{graphicx}
\usepackage[figuresright]{rotating}

\newcommand{\AmS}{{\protect\the\textfont2
  A\kern-.1667em\lower.5ex\hbox{M}\kern-.125emS}}

\title{3-d lattice SU(3) free energy to four loops \thanks{Presented by Andrea Mantovi}
\vskip-3.6cm\hfill\small MIT-CTP-3541; UPRF-2004-18\vskip3.3cm}
\author{F. Di Renzo \address[MCSD]{Dipartimento di Fisica, Universit\` a degli Studi di Parma, \\
        and INFN, Gruppo Collegato di Parma, Italia},
        A. Mantovi\addressmark[MCSD],
        V. Miccio\addressmark[MCSD],
        Y. Schroder\addressmark[]\address{Center for Theoretical Physics, MIT, Cambridge, MA, USA},         
        C. Torrero\addressmark[MCSD]}
       
\begin{document}
\begin{abstract}
We report on the perturbative computation of the 3d lattice Yang-Mills free energy 
to four loops by means of Numerical Stochastic Perturbation Theory.
The known first and second orders have been correctly reproduced; the third 
and fourth order coefficients are new results and the known logarithmic IR divergence 
in the fourth order has been correctly identified. Progress is being made in switching 
to the gluon mass IR regularization and the related inclusion of the Faddeev-Popov determinant.
\vspace{1pc}
\end{abstract}

\maketitle

\section{Introduction}

The main goal of finite temperature QCD is to characterize
the deconfinement transition bet\-ween the low temperature regime, ruled by
confinement, and the high temperature regime ruled by asymptotic freedom.
In this respect the free energy density, i.e. the $pressure$, is a sort of 
`theoretical laboratory' in which to study such a transition, and, 
eventually, a potential candidate observable in heavy ion collisions. 
Evidently, the large $T$ limit of the pressure is that of an ideal gas 
of non-interacting particles, $p \propto T^4$; 
unfortunately, lattice simulations across the whole temperature range 
are not feasible and our efforts and understanding are confined 
to the opposite ends of the temperature range: namely, the low temperature regime  
(up to some $4\div5$ times the transition temperature $T_c \sim 200$\,MeV)
which is accessible to actual computational resources,
and the high temperature regime at which the pressure exhibits 
a purely perturbative behaviour.

Dimensional reduction~\cite{BraNie} has been applied~\cite{York1} to fill the gap: 
$4d$ QCD is matched to $3d$ $SU(3)$ Yang-Mills (YM) coupled to a Higgs field in the adjoint
representation; such a theory can then be matched to 
$3d$ $SU(3)$ YM, which captures the ultrasoft degrees of freedom;
both these reductions have been successfully performed in the
$\overline{MS}$ scheme. Finally, $3d$ YM has to be treated non-perturbatively, 
the only effective means being the lattice. Indeed lattice computations  
can be consistently incorporated into the dimensional reduction strategy:
the theory is superrenormalizable and $all$ divergences can be computed
perturbatively; this allows a clean matching of the schemes in the
continuum. It is well known that computing at high orders in LPT 
is quite hard a task; in this respect the approach of 
Numerical Stochastic Perturbation Theory
(NSPT)~\cite{NSPT} is a very efficient one.

Let $f \equiv -V \ln Z$ be the free energy associated with the Wilson action 
$S_W = \beta_0 \sum_P (1 - \Pi_P)$,
$\beta_0=2N_c/(a^{4-d} g_0^2)$ being the dimensionless bare 
lattice inverse coupling in $d$ dimensions.
To compute the free energy one can revert to the computation of
the plaquette, being
\begin{equation} 
\langle 1 - \Pi_P \rangle = \frac{2\, a^d}{d(d-1)} \frac{\partial}{\partial \beta_0} 
\bigg(\frac{1}{T} f\bigg) \,;
\end{equation}
then, a weak-coupling expansion of the plaquette
\begin{eqnarray}
\langle 1 - \Pi_P \rangle &=&\frac{c_1(N_c,d)}{\beta_0} + \frac{c_2(N_c,d)}{\beta_0^2} \\
          &+&\frac{c_3(N_c,d)}{\beta_0^3} + \frac{c_4(N_c,d)}{\beta_0^4} + \cdots \nonumber
\label{PLexpns}
\end{eqnarray}
uniquely determines the corresponding expansion for $f$,
and for $N_c=3$ and $d=3$
\begin{eqnarray}
\frac{2}{6}\, \frac{1}{T} f = a^{-3} \left( c_0 + c_1 \ln \beta_0 \right) 
- a^{-2} \frac{c_2}{6} g_0^2 \\
- a^{-1} \frac{c_3}{72} g_0^4 - \frac{\tilde{c}_4}{648} g_0^6 + \emph{O}(a) \,. \nonumber
\label{PLtld}
\end{eqnarray}
so that the first four coefficients uniquely determine the matching 
to the continuum; $c_1$ and $c_2$ being long known, $c_3$ and $c_4$
were the aims of the computation. The presence has already been established~\cite{York2}
of a logarithmic IR divergence at fourth loop order; it was our aim to recover 
the scheme independent coefficient of such a logarithm:
the tilde on $c_4$ reminds us that the IR divergence has to be isolated 
and subtracted by means of a definite $IR$ $regulator$, which we choose 
to be the $finite$ $lattice$ $size$ $L$ inherent to any lattice simulation.

\section{Computational setup} \label{sec:Setup}

We have inserted the expansion
$U_\mu(n) = 1 + \sum_{i=1}^{8} \, \beta_0^{-\frac{i}{2}} \, U_\mu^{(i)}(n) \,
\label{Uexpns}$
into the Langevin equation
\begin{equation}
	\partial_tU_\eta=[-i\nabla S[U_\eta]-i\eta]U_\eta \,,
\end{equation}
which implements the strategy of Stochastic Quantization 
on lattice gauge fields ($\eta$ is a gaussian noise). 
We have integrated such an equation in the Euler scheme 
and linearly extrapolated to vanishing stochastic time step.
We have computed the coefficients of the expansion (3) 
on lattice sizes ranging from $L=5$ to $L=16$; up to three loops 
we also performed the computations on a $L=18$ lattice.

We finally extrapolated to infinite lattice size according to
the analysis by Symanzik which entails $subleading$ logarithms,
suppressed by inverse powers of $L$; recall that the finite volume 
acts as the IR regulator needed at 4 loops and enters a $leading$ logarithm. 
The indeterminacies we quote are dominated by this extrapolation procedure:
our data do not allow a clear cut between a logarithm and a constant and
we have been forced to effective extrapolations based on pure power-like fits;
the spreading of the results (figures) corresponds to different set of powers
included in the fit. More details can be found in~\cite{JHEP}.

Most of the computer simulations have been performed on a PC cluster
of 10 bi-processors Athlon MP2200. A programming environment for LGT--NSPT 
has been set up in C++ by V. Miccio with massive use of
(C++ specific) \emph{classes} and \emph{methods}
well suited to handle lattice and algebraic structures.
Some statistics came from a 14 Intel Xeon 2.0 Ghz PC cluster
which has become available along the way.

\section{Results}

In ~\cite{JHEP} we present the results we obtained for the coefficients $c_i^{(L)}$ 
at various $L$ and the values extrapolated to infinite volume;
the two benchmark coefficient are shown to be in excellent
agreement with the known results. Here we simply plot (figures 1,2) 
the extrapolating fits for the third and fourth coefficients, 
the new results, and quote the numbers:
\[
c_3 = -6.90_{(12)}^{(2)} \; , \; \; \; c_4 = -25.8(4) \; .
\]

\begin{figure}
\includegraphics[width=18pc]{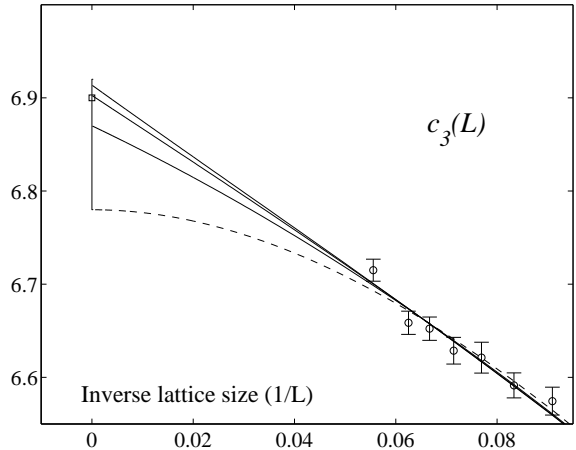}
\vspace{-1.2 cm}
\caption{Extrapolating fits for $c_3$.}
\end{figure}

In~\cite{JHEP} we discuss at length about the IR divergence at four-loop order;
let us just recall that it is quite hard to recognize the presence of a logarithm 
out of any set of data; we found that fits with no log
had slightly but systematically greater $\chi^2$ than fits with a
logarithmic divergence, and, most of all, that we could fit
quite well ($c_4^{(ln)}=1.1(2)$) the coefficient $c_4^{(ln)} =
81\,(688-157\pi^2/4)/(4\pi)^4 = 0.9765$ which has been computed in
~\cite{York2}.

\begin{figure}
\includegraphics[width=18pc]{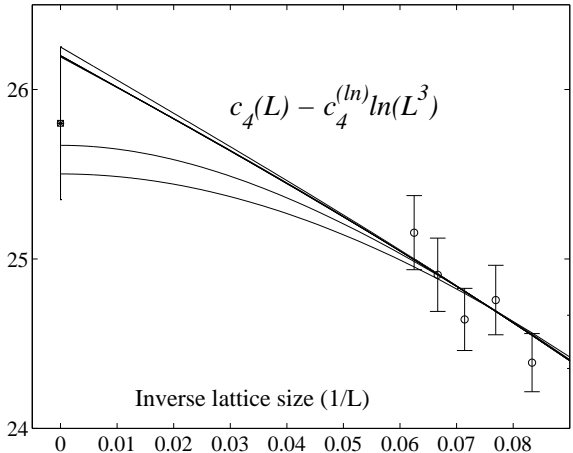}
\vspace{-1,2 cm}
\caption{Extrapolating fits for the log-subtracted $c_4$.}
\end{figure}

\section{Perspectives}

A definite IR regulator enables one to isolate a finite part 
out of the 4 loops contribution to the plaquette; we have been employing
the $finite$ $volume$ inherent to any computer simulation.
A $mass$ is a commonly used regulator in continuum computations; 
the $coupling$ itself (dimensionful in $3d$ YM) is quite a natural IR regulator
for nonperturbative lattice simulations.
Despite the fact that the coefficient of the logarithmic divergence is universal
it is clear that definite coefficients relate different regularizations.
We are interested in the perturbative matching between the lattice 
and a continuum perturbative scheme: in both approaches
the same gluon mass IR regulator can be employed 
so that the same mismatch shows up with respect to data coming 
from nonperturbative computer simulations with the coupling  
as IR regulator. Employing massive gluons 
is the na- tural extension of our computational programme.

Inserting a mass into the gluon propagator amounts to switching to the Lie
algebra for the definition of the new lattice action and this we
perform at a definite step of the computation. 
It is well known that the Lie algebra YM action is made out 
of a gauge fixing term and the ghost action resulting from the FP determinant.
Our NSPT strategy entails the construction of the Euler scheme solution 
$U' = exp(-\epsilon F) U$ to Eq.~(4); the quantity $F$ belongs to the Lie algebra 
and our algorithm entails going back and forth from the group to the algebra: 
at that stage we include the new contribution to the action.
The strategy for a determinant is well established~\cite{Cornell,FP} and is employed
by the Parma group in the unquenched NSPT approach to lattice QCD~\cite{UnquNSPT};
the determinant is simulated without introducing fermion 
or ghost fields and the inversions of the nonlocal matrix 
is performed perturbatively via an efficient FFT by a back and forth procedure
from Fourier space: the zeroth order of any propagator is diagonal in momentum space
and the construction of successive orders does not require any matrix inversion. 
The implementation of such a strategy is in progress;
signals of the stabilization of gauge dependent quantities
as well as of the invariance of gauge invariant quantities
make us confident about the reliability of the whole procedure.

\end{document}